\documentclass{aa}
\usepackage{graphicx}
\usepackage{txfonts}
\begin{document}
\title{On magnetic instabilities and dynamo action in stellar radiation zones}
\titlerunning{MHD instabilities and dynamo in stellar radiation zones}

\author{J.-P. Zahn\inst{1}, A. S. Brun\inst{2,1} \and S. Mathis\inst{3,2,1} }
\authorrunning{Zahn, Brun \and Mathis }

\offprints{A. S. Brun}

\institute
{LUTH, Observatoire de Paris, CNRS, Universit\'e Paris-Diderot; Place Jules Janssen, F-92195 Meudon, France\\
\email{jean-paul.zahn@obspm.fr} \and
DSM/DAPNIA/Service d'Astrophysique, CEA Saclay, F-91191 Gif-sur-Yvette, France; AIM, UMR 7158, CEA - CNRS - Universit\'e Paris 7 \and
Observatoire de Gen\`eve, 51 chemin des maillettes, CH-1290 Sauverny, Switzerland\\
\email{sacha.brun@cea.fr; stephane.mathis@cea.fr} \\
}

\date{Received 17 April 2007; accepted 18 July 2007}

\abstract
{We examine the MHD instabilities arising in the radiation zone of a differentially rotating star, in which a poloidal field of fossil origin is sheared into a toroidal field.}
{We focus on the non-axisymmetric instability that affects the toroidal magnetic field in a rotating star, which was first studied by Pitts and Tayler in the non-dissipative limit. If such an instability were able to mix the stellar material, it could have an impact on the evolution of the star. According to Spruit, it could also drive a dynamo.}
{We compare the numerical solutions built with the 3-dimensional ASH code with the predictions drawn from an analytical study of the Pitts \& Tayler instability.}
{The Pitts \& Tayler instability is manifestly present in our simulations, with its conspicuous $m=1$ dependence in azimuth. But its analytic treatment used so far is too simplified to be applied to the real stellar situation. Although the instability generated field reaches an energy comparable to that of the mean poloidal field, that field seems unaffected by the instability: it undergoes Ohmic decline, and is neither eroded nor regenerated by the instability. The toroidal field is produced by shearing the poloidal field and it draws its energy from the differential rotation. The small scale motions behave as Alfv\'en waves; they cause negligible eddy-diffusivity and contribute little to the net transport of angular momentum. }
{In our simulations we observe no sign of dynamo action, of either mean field or fluctuation type, up to a magnetic Reynolds number of $10^5$. However the Pitts \& Tayler instability is sustained as long as the differential rotation acting on the poloidal field is able to generate a toroidal field of sufficient strength. But in the Sun such a poloidal field of fossil origin is ruled out by the nearly uniform rotation of the deep interior. }

%
\maketitle

\section{Introduction}
In recent years there has been a surge of interest for stellar magnetism, due mainly to the discovery of magnetic fields in an increasing number of stars, and to their mapping through Zeeman imaging (cf. Donati et al. 2006). Theory benefits enormously from these new constraints, and quite naturally the main focus is on the generation of magnetic fields in turbulent convection zones, which can now been studied through high resolution numerical simulations (Brun et al. 2004, 2005; Dobler et al. 2006). But attention has been drawn also to the instabilities that may affect the magnetic field in stably stratified radiation zones. Spruit (1999) reviewed various types of instabilities that are likely to intervene in a magnetized radiation zone, and he concluded that the strongest among them were those which had been described by Tayler and his collaborators. Indeed, Markey and Tayler (1973) have shown that a purely poloidal field would be unstable to non-axisymmetric perturbations, 
and so would also a toroidal field (Tayler 1973; Wright 1973; Goossens et al. 1981). Later Pitts and Tayler (1985) proved that  even in the presence of rotation a toroidal field would be unstable to such perturbations. 

Spruit (1999, 2002) analyzed the latter instability in more detail, including Ohmic dissipation and radiative damping, and starting from the dispersion relation established by Acheson (1978); he suggested that it could regenerate the toroidal field, and thus drive a genuine dynamo. Applying essentially Spruit's prescriptions, Maeder and Meynet (2003, 2004, 2005), Eggenberger et al. (2005) and Heger et al. (2005) introduced this dynamo, and the turbulent transport believed to be associated with it, in their stellar evolution calculations. It proved extremely efficient in establishing quasi uniform rotation. In the case of massive stars, it would increase the loss of angular momentum through the stellar wind, and lead to slower rotating stellar cores; this would yield pulsar rotation rates that are in better agreement with the observations. 

However, guided by other observational evidence, Denissenkov and Pinsonneault (2006) questioned the existence of such a powerful process of angular momentum transport; they noticed some inconsistencies in Spruit's heuristic argumentation, which they tried to correct. Unfortunately the choice they made was in conflict with the rigorous analytical treatment of the instability; this was pointed out by Spruit (2006), and that part of their discussion was deleted in the final version (Denissenkov \& Pinsonneault 2007).  In the meanwhile, Braithwaite (2006) conducted a numerical experiment which he claimed to validate Spruit's dynamo scenario.

Our interest in those instabilities was aroused when we observed them while verifying if a fossil field was able to prevent the spread of the solar tachocline, as had been proposed by Gough and McIntyre (1998): the results of these three-dimensional simulations are reported in Brun and Zahn (2006). We then wanted to check whether these calculations agreed with Spruit's analytical predictions, and this incited us to re-examine his original derivation.

We begin by extending Acheson's dispersion relation to the case where both an entropy gradient and a composition gradient are present, and derive the actual solutions of that equation (rather than upper or lower bounds). We then discuss the heuristic arguments that were first used by Spruit, and show how they can be misleading. Next we compare the analytical solutions with the numerical results obtained in  Brun \& Zahn (2006, hereafter referred to as BZ06), and finally we examine whether the dynamo that was suggested by Spruit does actually operate.

\section{The non-dissipative case}
\label{non-diss}
Let us first recall the major results concerning the instability that affects a toroidal field in the radiation zone of a rotating star, which were derived by Pitts \& Tayler (1985) in the non-dissipative limit. We consider this field as given, and ignore what causes it, namely the mean poloidal field and the differential rotation that is acting on it; we further assume that it varies as $B_\varphi \propto \varpi^p$ with the distance from the rotation axis ($p \geq 1$). To ease the task for the reader, we put the technical developments in Appendix.

As was done in previous works, we perform a local analysis and examine the stability of an imposed axisymmetric toroidal field. We submit it to wave-like perturbations of the form
\begin{equation}
\exp i(l \varpi+m\varphi+nz-\sigma t),
\end{equation}
in a cylindrical reference frame $(\varpi, \varphi, z)$ centered on the rotation axis of a star that is rotating uniformly with angular velocity $\Omega$; $\sigma=\sigma_R + i \sigma_I$ is the complex frequency.
Such perturbations obey a dispersion relation that was first derived by Acheson (1978, Eq. 3.20); Spruit (1999) made it more tractable by applying it only to the vicinity of the rotation axis, by assuming that $(l/n) \ll 1$ and that the Alfv\'en velocity is negligible compared to the sound speed. In Appendix, we establish that equation in a slightly more general case, allowing the stratification to depend both on entropy and on chemical composition.
For $m \neq 0$ and ignoring all forms of energy dissipation (radiative, Ohmic, viscous) the dispersion relation takes the form (cf. Spruit 1999, Eq. A9)
\begin{eqnarray}
(\sigma^2 - \omega_A^2)\, ( \sigma^2 \!&-&\! \omega_A^2 [1+A ]) + (p-1) {\omega_A^2\over m^2} \left(2m \Omega \sigma+ 2\sigma^2\right) \nonumber \\
&-& \left(2\Omega \sigma+ {p+1\over m} \omega_A^2 \right) \left(2 \Omega \sigma+ 2{\omega_A^2 \over m} \right) =0 \,.
\label{non-dissipative}
\end{eqnarray}
We have introduced here the stratification parameter
\begin{equation}
A={l^2 \over n^2}  {N^2 \over \omega_A^2}
\end{equation}
which measures the relative strength of the two restoring forces: buoyancy and magnetic tension;
$N$ is the buoyancy frequency (cf. \ref{bv2}) and $\omega_A$ the Alfv\'en frequency associated with the toroidal field $B_\varphi$:
\begin{equation}
\omega_A^2 = \left({m V_A \over \varpi}\right)^2 = {m^2 \over \varpi^2} {B_\varphi^2 \over 4 \pi \rho} .
\end{equation} 

This dispersion equation admits two pairs of wave modes, that are well separated in frequency when $\omega_A \ll \Omega$, which is the case we shall consider here of  moderate field strength ($B_\varphi \ll100$ kG below the solar convection zone).
The fast modes may be isolated by setting $\sigma^2 \gg \omega_A^2$, which yields to first approximation $\sigma^2=(l/n)^2 N^2 + 4 \Omega^2$: we recognize here the signature of the gravito-inertial waves. The slow modes are found in the domain $\sigma^2 \ll \omega_A^2 \ll \Omega^2$, where the dispersion relation reduces to the following quadratic equation
in $\zeta=2 m\Omega \sigma/ \omega_A^2$
\begin{equation}
\zeta^2+4\zeta-[m^2 (1+A)-2(p+1)]=0 .
\end{equation}
Its discriminant is $\Delta ' = m^2 (1+A)+2(1-p)$;  when $\Delta ' < 0$, it has a complex root with a positive imaginary part $\sigma_I > 0$ and hence this mode would be unstable for
\begin{equation}
p > 1 + {m^2 \over 2} (1+A) .
\end{equation}
As Pitts and Tayler (1985) pointed out, all modes are stable for $p=1$. And one expects actually the toroidal field to vary linearly with $\varpi$ in the vicinity of the rotation axis, if it is generated by shearing a poloidal field through a depth dependent rotation. Hence we shall assume from here on that $p=1$.  We are then left with two oscillatory modes, whose frequencies are given by
\begin{equation}
\zeta = -2 \pm  m \, \sqrt{1+A}  \quad \hbox{or} \quad
\sigma= {\omega_A^2 \over \Omega} \left[-{1 \over m} \pm {\sqrt{1+A} \over 2} \right] .
\label{slow-modes}
\end{equation}
Note that for $A <3$ there are two negative roots, and that for $A>3$ the roots have opposite sign. 

\section{Overstable modes}
\label{sec-3}
Since all modes are stable for $p=1$ in the non-dissipative limit, instability can only occur in form of a diffusive instability, with the complex frequency presenting a positive imaginary part. The dissipation is provided here by radiative and Ohmic diffusion, with respective diffusivities $\kappa$ and $\eta$. In addition to the two slow oscillatory modes discussed above, whose amplitude may then grow exponentially in time, there is also a direct mode, whose frequency $\sigma_R  \rightarrow 0$ in the non-dissipative limit, that may become unstable.

We consider here only the case where the Roberts number $\varepsilon=\eta/\kappa$ may be treated as a small quantity, which is the case in stellar interiors ($\varepsilon \approx 10^{-4}$ below the solar convection zone). In Appendix, we derive Acheson's dispersion relation in the more general case where both a gradient of entropy and of chemical composition are present. Then the stratification is characterized by two buoyancy frequencies:
\begin{equation}
N^2 = N_t^2 + N_\mu^2  = {g  \over H_P} (\nabla_{\rm ad} - \nabla)
+   {g \over H_P} \left({d \ln \mu \over d \ln P}\right) 
\label{bv2}
\end{equation}
with the usual notations, and assuming the perfect gas law for simplicity.
The corresponding stratification parameters are then
\begin{equation}
A_t = {l^2 \over n^2} {N_t^2 \over \omega_A^2} \quad {\rm and } \;
A_\mu= {l^2 \over n^2} {N_\mu^2 \over \omega^2_A} ;
\end{equation}
it is convenient to introduce also the sum  $A^* = \varepsilon A_t + A_\mu$.

In Appendix (Eq. \ref{disp-reduced})
we show that the scaled complex frequency $\zeta  = 2 m \Omega \, \sigma / \omega_A^2$ obeys the following third-order equation: 
 \begin{eqnarray}
\varepsilon A_t  \zeta^2 &+& (2 \zeta^2+4\zeta-A_\mu) \, S^2  \nonumber \\
& -& i\, S \,\zeta \,   (\zeta^2 + 4\zeta + 3-A^* -S^2) =0 ,
\label{disp-reduced1}
\end{eqnarray}
where  $\, S = 2 \Omega \eta n^2 / \omega_A^2$ represents the suitably scaled Ohmic diffusivity.  Note that $A_t$ intervenes only through the product $\varepsilon A_t$, which illustrates how the restoring force due to the entropy gradient is weakened through radiative damping. For given meridional wavenumbers $n$ and~$l$, or equivalently for a set $[A_t$, $A_\mu$, $S]$, Eq.~\ref{disp-reduced1} has three complex solutions; if its imaginary part $\zeta_I$ is positive, the mode is overstable. Its growth-rate $\sigma_I= (\omega_A^2/2m\Omega)\, \zeta_I$  then peaks at azimuthal wavenumber $|m|=1$, a distinctive property of this instability. From here on we shall consider only that case $|m|=1$, and refer to this instability as the Pitts \& Tayler instability, to distinguish it from that which affects a poloidal field (Markey \& Tayler 1973).

 We observe that if $S$, $\varepsilon A_t$ and/or $A_\mu$, are of order unity, this will also be the case for $\zeta$, which implies that  both the frequency $\sigma_R$ and the growth-rate $\sigma_I$ scale as $\omega_A^2/\Omega$, like in the non-dissipative case. The maximum growth-rate and the corresponding frequency are given in Table~\ref{table-max} for a sample of values of $\varepsilon A_t/A_\mu$;
we see that it is the prograde mode that is unstable (or rather overstable), except for $A_\mu=0$.

For given ratio $A_t/A_\mu$, the instability domain can be delineated in the $[A^*, S]$ plane by locating the points where the growth-rate vanishes and thus where the oscillation frequency is purely real: $\sigma=\sigma_R$.  Referring back to (\ref{disp-reduced1}), this occurs when the following equations  are both satisfied:
\begin{eqnarray}
\varepsilon A_t \, x^2+  (2 x^2+4x-A_\mu)\, S^2&=& 0 \, ,  \nonumber  \\  
x^2 + 4x + 3-A^* - S^2&=&0 \, , \label{coupled}
\end{eqnarray}
where $x=2m \Omega \, \sigma_R/\omega_A^2$ is the scaled frequency. The solution of this system determines the minimum strength of the magnetic field that is required for instability, as we shall see next. 

Let us first examine the two limit cases ($A_t=0$ and $A_\mu=0$) that were considered by Spruit.

\subsection{Stratification due only to the composition gradient}
\begin{figure}
  \centering
  \includegraphics[width=8cm]{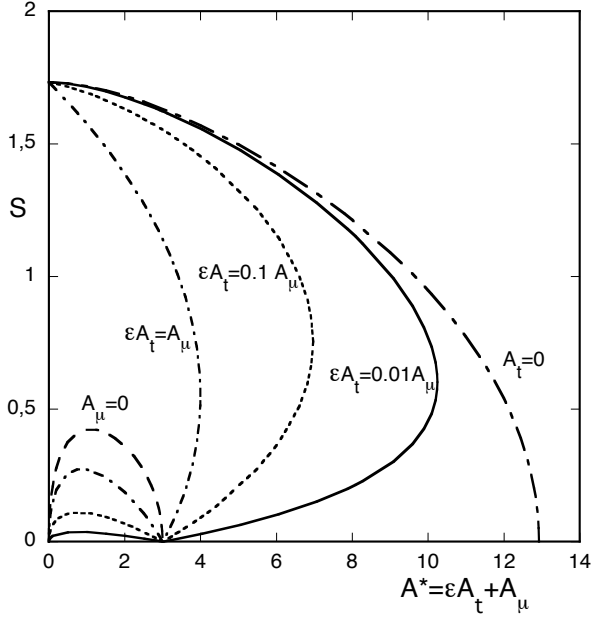}
      \caption{Limits of the instability domain in the $(A^*, S)$ plane, for various values of $A_t/A_\mu$. The stratification parameters are defined as $A_t=(l/n)^2 (N_t/\omega_A)^2$ and $A_\mu=(l/n)^2 (N_\mu/\omega_A)^2$, whereas $A^*=\varepsilon A_t+ A_\mu$. $\varepsilon=\eta/\kappa$ designates the Roberts number, and $S=2 \eta \Omega n^2/\omega_A^2$ defines the limits of the instability domain in vertical wavenumber $n$, for given $A^*$.
              }
         \label{fig.1}
   \end{figure}

For $A_t=0$, system (\ref{coupled}) reduces to 
\begin{eqnarray}
2 x^2+4x-A_\mu\, &=& 0  \, ,  \nonumber  \\  
x^2 + 4x + 3-A_\mu - S^2&=&0 \, .  \label{coupled-amu}
\end{eqnarray}
It is easily solved:
\begin{eqnarray}
x&=&-1+\sqrt{1+A_\mu/2} \nonumber \\
S^2 &=& 2  \sqrt{1+A_\mu/2} +1 - A_\mu/2 ;
\end{eqnarray}
the function $S(A_\mu)$ decreases from $\sqrt{3}$ at $A_\mu=0$ to $0$ at $A_{\rm max} =6+4\sqrt{3}=12.928$, as illustrated in Fig.~\ref{fig.1}.

We now use this result to determine the instability threshold. Like Spruit (1999), we eliminate $n^2$ by combining the definition of $A_\mu$:
\begin{equation}
{l^2 \over n^2} \, {N_\mu^2 \over \omega_A^2}= A_\mu 
\label{amu}
\end{equation}
with that of $S$:
\begin{equation}
 {2 \Omega \eta \over  \omega_A^2}  \, n^2 = S(A_\mu) ;
\end{equation}
this gives us  the minimum strength of the toroidal field required to trigger the instability:
\begin{equation}
[A_\mu\, S(A_\mu)] \, \omega_A^4 = 2 \Omega \eta l^2 N_\mu^2 .
\label{threshold-eta}
\end{equation}
That condition is optimized when $[A_\mu \, S(A_\mu)]$ is maximum, at $[AS]_{\rm max}=9.829$, which is reached for $A_\mu=8.788$.
Thus our result confirms that of Spruit (1999, Eq. A23), although they differ somewhat in their numerical coefficients ($A_{\rm max}$, for instance, is underestimated by Spruit, whose value is 3 instead of 12.928).

\subsection{Stratification due only to the entropy gradient} 
For $A_\mu = 0$ system (\ref{coupled}) has no simple analytical solution, but it can be easily shown that $S(A_t)$ vanishes at $\varepsilon A_t=0$ and $3$, as illustrated in Fig.~\ref{fig.1}. Thus the maximum value of the stratification parameter  that allows for instability is $A_t = 3 /\varepsilon$.

Proceeding as before, we find that the instability threshold is given by 
\begin{equation}
[\varepsilon A_t\, S(\varepsilon A_t)] \, \omega_A^4 = 2 \Omega \varepsilon \eta l^2 N_t^2 ;
\end{equation}
it is optimized at $[\varepsilon A_t\, S(\varepsilon A_t)]=0.728$ for $\varepsilon A_t = 2.070$. Here also we retrieve Spruit's result, but with the exact proportionality constant.

\begin{table}[htdp]
\caption{Maximum scaled growth-rate $\zeta_I^{\rm max}=2 \Omega \sigma_I/\omega_a^2$ and corresponding frequency $\zeta_R^{\rm max}=2 \Omega \sigma_R/\omega_a^2$, $A^*$ and $S$,  at given $\varepsilon A_t/A_\mu$ and for $|m|=1$.}
\label{table-max}
\centering
\begin{tabular}{c c c c c}
\hline\hline
$\varepsilon A_t  / A_\mu$ &  $A^*$ & $S(A^*)$ & $\zeta_R^{\rm max}$ & $\zeta_I^{\rm max}$ \\
\hline
0 & 3.306 & 0.3463 & 0.3868 & 0.1722 \\
0.01 &3.203 & 0.3593 & 0.3736 & 0.1679 \\
0.1 & 2.957 & 0.3774 & 0.2820 & 0.1397 \\
1 & 1.817 & 0.3750 & 0.1160 & 0.06945 \\
10 & 1.908 & 0.2599 & 0.02211 & 0.01736 \\
$\infty$ & 0 & 0 & -- 0.4858 & 0.4227 \\
\hline
\end{tabular}
\end{table}
\begin{table}[htdp]
\caption{Stratification parameters and optimal conditions for the onset of instability. $A^*_{\rm max}$ is the highest value of $A^*=\varepsilon A_t+A_\mu$ that allows for instability; $[AS]_{\rm max}$ the maximum of the product $A^*S(A^*)$ for fixed $\varepsilon A_t/A_\mu$, obtained at $A^*_{\rm opt}$, from which one derives the instability threshold for the magnetic field (see text).}
\label{table1}
\centering
\begin{tabular}{c c c c c}
\hline\hline
$\varepsilon A_t  / A_\mu$ &  $A^*_{\rm max}$ & $S(A^*_{\rm max})$ &$[AS]_{\rm max}$ & $A^*_{\rm opt}$ \\
\hline
0 & 12.928& 0 &9.829 & 8.788 \\
0.01 &10.249 & 0.602 &9.307 & 8.138 \\
0.1 & 6.970 & 0 .754 &6.960 & 5.885 \\
1 & 4.000 &0.550& 3.451 & 3.171 \\
10 & 3.125 & 0.219 & 2.186 & 2.087 \\
$\infty$ & 3& 0 & 0.728 & 2.070 \\
\hline
\end{tabular}
\end{table}

\subsection{Stratification due to both entropy and composition gradients}
In the general case both causes of stratification are present, and $S$ is a function of  $ A_t$ and $A_\mu$; the instability domain is displayed in Fig.~\ref{fig.1} for a set of fixed $A_t/A_\mu$. All solutions, except for $A_\mu=0$, have in common the point $A^*=0$, $S=\sqrt{3}$, and all solutions, except for $A_t=0$ share the point $A^*=3, \; S=0$, as predicted by (\ref{coupled}). Note that they are double-valued in the general case, which means that the instability interval in $n$, the vertical wavenumber, has a lower boundary that differs from zero. Only in the limit cases, $A_t=0, A_\mu=0$, does the instability domain extend to $S=0$, between $A^*=0$ and $A^*_{\rm max}$.

To determine the instability threshold for the toroidal field, we eliminate as before $n^2$ between
\begin{equation}
{l^2 \over n^2} \left[ \varepsilon {N_t^2 \over \omega_A^2}  +{N_\mu^2 \over \omega_A^2}\right] = A^*
\label{astar}
\end{equation}
and
\begin{equation}
 {2 \Omega \eta \over  \omega_A^2}  \, n^2 = S(A^*) ;
\end{equation}
this yields 
\begin{equation}
[A^*\, S(A^*) ]\, \omega_A^4 = 2 \Omega \eta l^2  \left[\varepsilon N_t^2  + N_\mu^2 \right] ,
\label{inst-crit-general}
\end{equation}
which validates Spruit's prescription of replacing the buoyancy frequency $N^2$ by $[\varepsilon N_t^2  + N_\mu^2]$ (his Eq. 55).\footnote{Through heuristic arguments, Maeder \& Meynet (2004) obtain a similar expression, but where $N_t^2$ is reduced by $\varepsilon/2$ instead of $\varepsilon$.}
This expression is optimized for the values of $[A^*\, S(A^*)]$ that are given in Table~\ref{table1}, together with the maximum value of the stratification parameter $A^*_{\rm max}$ which allows for instability, and that of the corresponding $S(A^*_{\rm max})$.

For either $\varepsilon=0$ or $N_\mu^2=0$, we retrieve the two special cases considered before.

\section{On the validity of heuristic arguments}

In his seminal paper, Spruit (1999) used a few heuristic arguments to describe the Pitts \& Tayler instability in a simple, intuitive  way. Let us examine their validity in the light of the rigorous treatment given in the preceding section and in Appendix. 

With one such arguments, Spruit seeks a lower limit for the vertical wavenumber $n$.  We quote him: ``For displacements of amplitude $\xi$, the work done per unit mass against the stable stratification is ${1\over 2} \xi^2 (l/n)^2 N^2$. The energy gained from the field configuration is ${1\over2}\omega_A^2 \xi^2$. For instability, the field must be strong enough, such that $\omega_A^2 > (l/n)^2 N^2$." Furthermore, since $l \simeq 1/r$, this provides him a lower limit for $n$ (cf. his Eq. 44). 

In Spruit (2002), the argument is slightly different: $\xi$  is the {\it unstable} displacement, and ${1\over2}\omega_A^2 \xi^2$ the kinetic energy that is released in this displacement, ``since the growth-rate of the instability in the absence of constraints like rotation is $\sigma \sim \omega_A$". 

Another argument is used to set an upper limit to $n$: the growth-rate $\sigma$ has to overcome the damping rate of Ohmic diffusion; hence 
\begin{equation}
\sigma > n^2 \eta \, ,
\label{eta-diffusion}
\end{equation}
and this is applied to the case $\Omega \gg \omega_A$, where $\sigma \sim \omega_A^2 / \Omega$.

Denissenkov \& Pinsonneault (2006)  object quite understandably to this ambivalent choice of $\sigma$, namely $\sigma\sim \omega_A$ for the lower bound and $\sigma\sim \omega_A^2 / \Omega$ for the upper bound on $n$. They take the latter definition in both cases, which seems more coherent - but disagrees with the rigorous treatment.

Where is the problem? In fact buoyancy and magnetic field act both as restoring forces, as is illustrated by the energy equation, which we establish in Appendix in the non-dissipative case  (\ref{energy}):
\begin{equation}
{1 \over 2} \left({\partial \xi \over \partial t}\right)^2  +
{1 \over 2}m^2 \omega_A^2  \xi^2 +  
{1 \over 2}N^2  \xi_z^2 =0 .
\label{energy2}
\end{equation}
It shows how the energy of the perturbation is exchanged back and forth between kinetic and potential (magnetic and buoyancy) energy, as is typical in dynamical systems. In the absence of dissipation, the total energy remains constant, while it increases when the diffusive instability occurs.
Thus the buoyancy force and the Lorentz force\footnote{In fact, one should call it the Laplace force, contrary to what has become common practice. Pierre-Simon Laplace (1749-1827) was the first to give the expression of the force exerted by a magnetic field on an electric current; the force named after Hendrik Lorentz (1853-1928) is that experienced by a moving charged particle.} are not antagonistic, and it is not possible to draw an instability condition from their relative strengths. 

These shortcomings of the heuristic approach have been implicitly recognized in Spruit (2006),  where he writes that ``all physical effects contribute if all terms and factors [meaning $A_\mu, \, \varepsilon A_t, \,S$ and $\zeta$ in our notation] are of the same order [i.e. of order unity, as can be seen in (\ref{disp-reduced1})]". To achieve this, the ratio $(l/n)^2$ has thus to satisfy $(l/n)^2 \sim \omega_A^2/N^2$, a condition that is less stringent than Spruit's original inequality.

The other heuristic condition (\ref{eta-diffusion}) deserves also a couple of remarks. First, as we show in Appendix, in the limit of small $n^2$ the growth-rate is proportional to the damping rate: $\sigma \propto n^2 \eta$ (cf. \ref{growth-amu}), and  it is not clear what information can be extracted then about $n^2$ by equating these two rates.  Second, in presence of radiative damping, why should  this condition not be replaced by $\sigma > n^2 \kappa$, since the thermal diffusivity $\kappa$ largely exceeds  the Ohmic diffusivity? 

We conclude that heuristic arguments, though appealing, can be misleading when they are not supported by a rigorous treatment.

\section{Can the Pitts \& Tayler instability sustain a dynamo?}

We now come to the  suggestion made by Spruit (2002),  that this instability could sustain a dynamo in stellar radiation zones,  as thermal convection does in a convection zone. We find his idea quite interesting, but we argue that this dynamo cannot operate as he describes it.
According to him, the instability-generated small scale field, which has zero average, is wound up by the differential rotation ``into a new contribution to the azimuthal field. This again is unstable, thus closing the dynamo loop." But this shear induced azimuthal field has the same azimuthal wavenumber as the instability-generated field,
i.e. $m \neq 0$ and predominantly $m =1$: it has no mean azimuthal component, and thus it cannot regenerate the mean toroidal field that is required to sustain the instability. 
 For the same reason,  the instability-generated field cannot regenerate the mean poloidal field, as was suggested by Braithwaite (2006). Therefore the Pitts \& Tayler instability cannot be the cause of a dynamo, {\it as it was described by Spruit and Braithwaite.}

\begin{figure}
  \centering
\includegraphics[width=\linewidth]{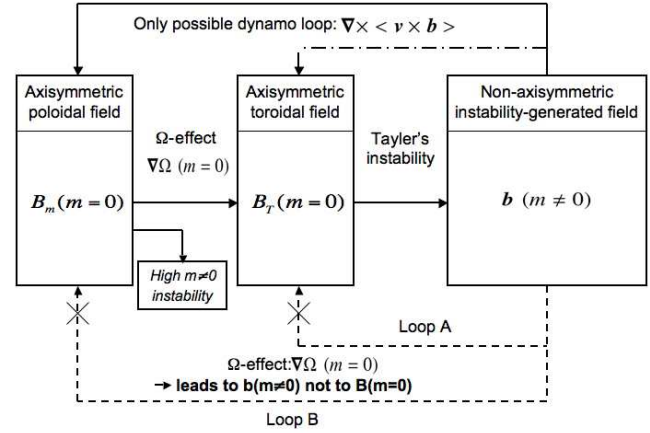}
      \caption{How to close the dynamo loop involving the Pitts \&Tayler instability. In dashed lines the loops proposed by Spruit (A) and Braithwaite (B). The only possible way to regenerate the mean toroidal and/or poloidal field is through the mean electromotive force $<\vec v \times \vec b>$ produced by the non-axisymmetric instability-generated field. But the dynamo must regenerate the poloidal field, in order to be fed from the differential rotation, and this leaves only the loop drawn in solid line. 
              }
         \label{fig.2}
   \end{figure}

To close the dynamo loop, as is well known in mean-field theory (Parker 1955; Moffatt 1978), one has to invoke the so-called $\alpha$-effect, which involves the non-zero mean electromotive force $<\vec v \times \vec b>$ that is produced here by the Pitts \& Tayler instability. This can be read in the azimuthally averaged induction equation:
\begin{equation}
{ d \vec B \over d t} = \vec e_\varphi \left[\varpi \vec{B_m} \cdot \vec \nabla \Omega \right] \, + \,
\vec \nabla \, \times <\vec v \times \vec b>  - \vec \nabla  \times (\eta \vec \nabla \times \vec B) ,
\end{equation}
where $\vec v$ and $\vec b$ are the non-axi-symmetric parts of the velocity and magnetic fields;
the meridional advection has been absorbed in the Lagrangian time derivative. 
The first term on the RHS describes how the poloidal field $\vec{B_m}$ is wound up by 
the differential rotation to produce the mean toroidal field (the $\Omega$-effect), the second how the mean electromotive force may (re)generate both the poloidal and the toroidal fields (the $\alpha$-effect), and the last term represents the Ohmic diffusion, eventually enhanced by the turbulence (the $\beta$-effect). The only possible dynamo loop is depicted in Fig.~\ref{fig.2} (solid lines); those proposed by Spruit and Braithwaite are shown in dashed lines. Moreover, for the dynamo to operate, the mean electromotive force must overcome the Ohmic dissipation.

 Another type of dynamo has been observed in numerical simulations, with externally forced turbulence: it is the {\it small-scale or fluctuation dynamo}. There the magnetic field has no mean components, neither poloidal nor toroidal, but a quasi-stationary regime may be achieved when the magnetic Reynolds number Rm=$VL/\eta$ exceeds some critical value ($V$ and $L$ are the characteristic velocity and length-scale of the turbulence).
But it remains to be checked whether such a dynamo can be sustained by an imposed shear, such as a differential rotation.

\section{3-D numerical simulations of the MHD instabilities}
\label{section-simus}
In Spruit's scenario, the toroidal field which undergoes the Pitts \& Tayler instability is generated by winding up an existing poloidal field through a differential rotation $\Omega(z)$ that results from slowing down the star by a stellar wind. It is similar to the model we took to examine the possibility of confining the solar tachocline by a fossil field (BZ06). However, in our case the differential rotation is imposed in latitude by the adjacent convection zone, and the depth dependence of $\Omega$ near the polar axis is caused by thermal diffusion
 (cf. Spiegel \& Zahn 1992).  What distinguishes our model from that of Spruit, and several others (cf. Miesch et al. 2007;  Arlt et al. 2007; Kitchatinov \& R\"udiger 2007), is the presence of that large scale poloidal field which is allowed to evolve freely through advection by the meridional circulation, Ohmic diffusion, and eventually interaction with the instability-generated field.
 \begin{table}
\begin{center}
\begin{tabular}{|*{5}{c|}} 
   \hline 
   Parameter & Symbol & Sun & Case A & Case B  \\ 
   \hline     
  thermal diffusivity & $\kappa$ & $10^7$ & $8 \, 10^{12}$ & $8 \, 10^{12}$ \\
    magnetic diffusivity & $\eta$ & $10^3$ & $8 \, 10^{10}$ & $8 \, 10^{9}$ \\
    viscosity & $\nu$ & $30$ & $8 \, 10^{9}$  & $8 \, 10^{9}$\\  
        buoyancy frequency & $N_t$ & $ 2.1 \,10^{-3}$ & $ 3 \, 10^{-4}$ &$ 3 \, 10^{-4}$\\
       rotation frequency & $\Omega$ & $3 \, 10^{-6}$ & $3 \, 10^{-6}$ & $3 \, 10^{-6}$\\
          \hline 
\end{tabular}
\bigskip 
  \caption{Typical values of the relevant parameters in the upper radiation zone of the Sun, and values adopted for the numerical simulations (in cgs units). \label{tablecoeff}} 
   \end{center} 
\end{table}

Our simulations, together with the equations of the problem and the resolution methods, are described in  full detail in BZ06.  We used the global ASH code (Clune et al. 1999; Brun et al. 2004) to solve the relevant anelastic MHD equations (Eqs. 1 - 5 in BZ06) in a spherical shell representing the upper part of the solar radiation zone ($0.35 \leq r/R_\odot \leq 0.70$), using a resolution of $N_r \times N_\theta \times N_\varphi= 193 \times 128 \times 256$.
For numerical reasons, we had to increase substantially the diffusivities of heat, magnetic field and momentum, as shown in Table~\ref{tablecoeff}, while respecting their hierarchy in the solar conditions. The characteristic evolution times are shortened accordingly, but contrary to BZ06 we made no effort here to rescale them by the Eddington-Sweet time in order to facilitate the comparison with the real Sun. In addition to the case A discussed in BZ06, we performed an additional series of simulations with a lower Ohmic diffusivity (by a factor of 10, case B), in order to reach a higher magnetic Reynolds number. 
\begin{figure}
   \centering
  \includegraphics[width=1.0\linewidth]{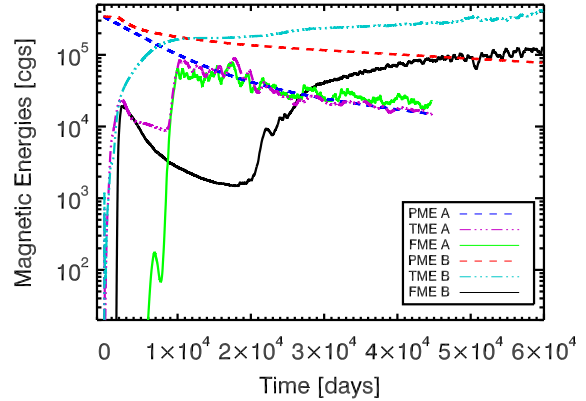}
      \caption{Time evolution of the energies of the mean poloidal (PME), mean toroidal (TME) and non-axisymmetric (FME) components of the magnetic field. Cases A and B  refer respectively to higher and lower magnetic diffusivity (cf. Table~\ref{tablecoeff}). Note the steady decline of the poloidal field, which is not affected by the irruption of the $m=1$  Pitts \& Tayler instability (at t$\approx$ 8,000 days in case A and $\approx$ 20,000 days in case B).
              }
         \label{fig.3}
   \end{figure}

\begin{figure}
   \centering
  \includegraphics[width=1.0\linewidth]{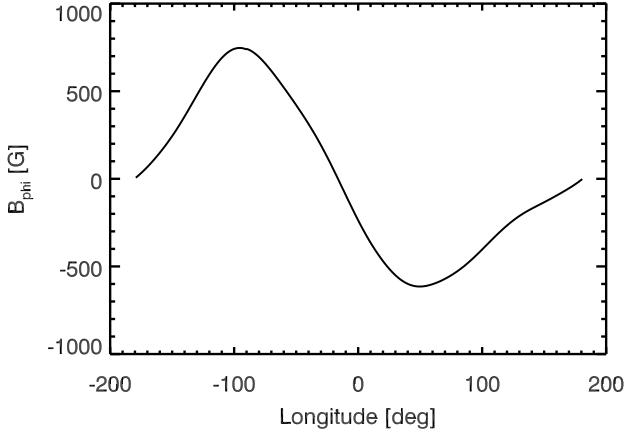}
      \caption{Azimuthal component of the instability-generated magnetic field, at the peak of the instability ($\approx$ 10,000 days, case A), measured at $r/R_\odot = 0.7, \, \theta = 75 ^o$. Note that this non-axisymmetric field is dominated by the wavenumber $m=1$, which is the signature of the Pitts \& Tayler instability.
              }
         \label{fig.4}
   \end{figure}

The temporal evolution of the magnetic fields and of the MHD instabilities is best followed in Fig.~\ref{fig.3}, where we display the energies of the poloidal, toroidal and non-axisymmetric components of the field.
Initially a purely poloidal field of about 1~kG (when measured at the base of the computational domain) is buried in the radiation zone; it is unstable to non-axisymmetric perturbations of high azimuthal wavenumber ($m \approx 40$), as shown in BZ06 (cf. their Fig.~7), which is in agreement with the predictions of Markey and Tayler (1973). 
The poloidal field diffuses outward at a rate proportional to the Ohmic diffusivity  and at some point (around 8,000 days in case A or 20,000 days in case B) it meets the differential rotation that has spread into the radiation zone, due to thermal diffusion. Their interaction induces there a toroidal field, whose strength is comparable to that of the poloidal field in case A; it becomes even larger in case B, where it keeps growing when we stop the simulation, at 60,000 days (corresponding to 1.3 Gyr when rescaled to the Ohmic diffusivity of the Sun).
This toroidal field produces a strong non-axisymmetric MHD instability, with a dominant azimuthal wavenumber $m=1$,  which is clearly the signature of the Pitts \& Tayler instability: it is illustrated in Fig.~\ref{fig.4}. In case A,  this instability-generated field saturates at an energy comparable to that of the mean poloidal field, whereas in case B it is still increasing when we stop the simulation, much like the toroidal field.

\begin{figure}
   \centering
  \includegraphics[width=1.3\linewidth]{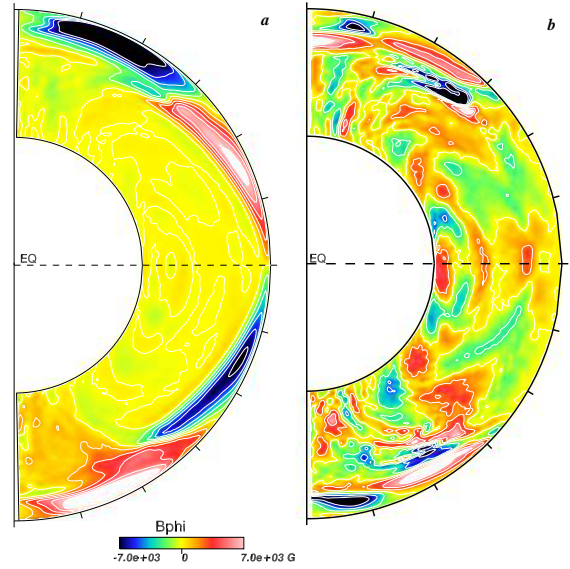}
      \caption{Mean toroidal field $B_\varphi(r, \theta)$ (left) and meridional section of the fluctuating toroidal field $b_\varphi (r,\theta, \varphi)$ (right). Only the odd $m$ have been kept in the latter, and the azimuthal angle $\varphi$ has been chosen such as to  emphasize the $m=1$ component. Note that mean and fluctuating fields are of comparable strength. (Case B, $t=50,000$ days.)
              }
         \label{fig.5}
   \end{figure}

This can be seen also in Fig.~\ref{fig.5}, where we display a meridional section of the mean and fluctuating components, from the case B simulation at $t=50,000$ days, where the energy of the fluctuating field matches that of the poloidal field. By subtracting the data of opposite azimuthal angles, $\varphi$ and $\varphi + \pi$, we isolate the modes of odd $m$, and we choose $\varphi$ such as to optimize the contribution of the $m=1$ component.
The mean field $B_\varphi$ has a quadrupolar shape and is strongest close to the convection zone, whereas the fluctuating field $b_\varphi$ is present everywhere, at medium level. At high latitude, the fluctuations are broader in the horizontal than in the vertical direction; translated into wavenumbers, this anisotropy amounts to $n/l \approx 5$. Measured in the vertical direction,
the characteristic scales are somewhat larger for $B_\varphi$ ($\approx 0.07 R_\odot$) than for $b_\varphi$ ($\approx 0.05 R_\odot$).

\section{Confronting the numerical simulations with the analytical model}
How do our 3-D simulations compare with the simplified analytical model described above in \S2-3?
The Alfv\'en frequency $\omega_A \sim 3\,10^{-7}$ s$^{-1}$ that characterizes the toroidal field at the peak of the instability ($\approx 10^4$ days in  case A) is significantly below the critical value for the Pitts \& Tayler instability which is deduced from (\ref{inst-crit-general}): $\omega_A^{\rm crit}= 2\,10^{-6} \, \sqrt{R/l_\varpi}$ s$^{-1}$, where $l_\varpi= 2 \pi/l$ is the radial wavelength. 
   
Another property of the analytical model is that the vertical wavenumber largely exceeds the horizontal wavenumber, particularly when the stratification is dominated by the $\mu$-gradient, since $(l/n)^2 (N_\mu / \omega_A)^2 = A_\mu={\cal O}(1)$ (Eq. \ref{amu}). In our model, the contrast $n/l$ is less severe, because there is no composition gradient: the instability threshold is then given by (\ref{astar}), which we rewrite here as
\begin{equation}
\varepsilon {l^2 \over n^2}  {N_t^2 \over \omega_A^2} = \varepsilon A_t \leq 3 .
\end{equation}
With the parameters that characterize our numerical simulations, this yields $n/l \gtrsim 50$, which is still not compatible with the observed structure of our $m=1$ mode, where $n/l \approx 5$. It is as if the role of the stratification were overestimated in the analytical model.

Therefore we must conclude that the analytical model considered by Spruit is not applicable to the more realistic problem treated in BZ06, and to the situation arising in stars. There may be several reasons for this. The first is that this simplified model ignores the presence of the poloidal field. However its inclusion would probably tend to stabilize the configuration, and that would worsen the discrepancy.

Second, the rotation is taken as uniform, whereas the shear of the differential rotation certainly plays a destabilizing role. Moreover, it is that neglected differential rotation which is the very cause of the instability by generating the toroidal field, a point that has been stressed by Maeder \& Meynet (2003).

The third reason is that the $\varpi$ dependence of the instability is poorly described by a trigonometric function, unless the wavelength is small compared to the distance to the axis: $l_\varpi \ll \varpi$. Bessel functions would probably be better suited (cf. Tayler 1957).

Finally the simplified model ignores the curvature effects, as was emphasized by Denissenkov \& Pinsonneault (2006). The neglect of the $\varpi$-component of the buoyancy force, compared to that of the Lorentz force, is allowed only as long as  $N^2 (\varpi/r) \,\xi_\varpi \ll \omega_A^2 \xi_\varpi$, hence in close vicinity of the rotation axis, for
$\varpi /r \ll (\omega_A/N)^2$. This is a very severe restriction, since $(\omega_A/N)^2 \sim 10^{-6}$ in our numerical solution (case A).
 
\section{Looking for dynamo action} \label{section-dynamo} 

An important property of our numerical solutions is that the decline of the poloidal field is not affected by the instability-generated field.
As can be seen in Fig.~\ref{fig.3}, this is true even once the (Pitts \& Tayler) instability has reached its saturation level, where its energy is comparable with that of the mean poloidal field. This has two consequences. First it proves that the smallest resolved scales do not act on the mean poloidal field as a turbulent diffusivity: they seem to behave rather as gravito-Alfv\'en waves, with their kinetic energy balancing their potential energy (magnetic + buoyancy),  as in the linear regime described by Eq. (\ref{energy2}). Hence the saturation of the instability is not due to the mechanism suggested by Spruit (2002), which was inspired by thermal convection,  namely that the magnetic eddy-diffusivity $\eta_t$ adjusts such as to neutralize the Pitts \& Tayler instability, i.e. that it satisfies eq.~\ref{inst-crit-general}. In our model - and presumably also in stellar radiation zones~- the regulation is apparently achieved through the action of the Lorentz torque on the differential rotation, which produces just the right amount of toroidal field that is required to sustain the instability, through small departures from Ferraro's law. 

At saturation, the mean quantities are stationary on the instability time-scale $\Omega/\omega_A^2$, which translates into the following condition for the azimuthal component of the momentum equation: 
\begin{equation}
4 \pi \rho {\partial \over \partial t} (\varpi^2 \Omega)= \vec{B_m} \cdot \vec\nabla (\varpi B_\varphi)\, + <\! \vec{b_m} \cdot \vec\nabla (\varpi b_\varphi)\!> \,
\approx  0 .
\end{equation}
Here $\vec B$ is the mean axisymmetric magnetic field with $\vec{B_m}$ being its meridional part, and  likewise for  $\vec b$,  the non-axisymmetric field generated by the instability; $< \, . \, . \, . \,>$ designates the azimuthal average. Since the characteristic scales are comparable for the mean and fluctuating field, this equation tells us that the magnetic energy of the instability field must be - crudely speaking - of the same order as that of the axisymmetric mean field, as we observe in our simulations.

 In our simulations we see no regeneration of the mean poloidal field: the $\alpha$-effect plays a negligible role,  at least up to the magnetic Reynolds number $Rm\,=\,R^2 \Delta \Omega / \eta \sim 10^5$ (case B), for Prandtl number $P_m=\nu/\eta=1$. (Evaluated with the instability velocity $\vec v$ at saturation, that Reynolds number is of the same order: Rm$\,=v_\varpi \xi_\varpi/\eta = (v_\varpi^2 /\sigma_I) /\eta \sim V_A^2 (\Omega/\omega_A^2)/\eta = \varpi^2 \Omega/ \eta$.) 
Note that the $\beta$-effect, i.e. the turbulence-enhanced diffusivity, is also absent here; hence one should not expect much mixing of the stellar material.
 We thus conclude that in our simulations the Pitts \& Tayler instability is unable to sustain a large-scale mean field dynamo, in the parameter domain that we have explored.

   \begin{figure}
   \centering
 \includegraphics[width=\linewidth]{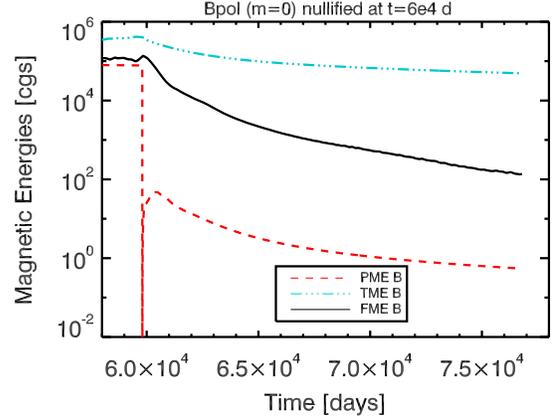}
          \caption{Evolution of the magnetic energies after suppressing the mean poloidal field  at $t=60,000$ days (case B). The mean electromotive force due to the instability-generated field produces some amount of poloidal energy (PME, dashed line), but that field is too weak to prevent through $\Omega$-efect the Ohmic dissipation of the toroidal field $B_\varphi$ (TME, dash-dotted line), which causes the decline of the instability-generated field (FME, solid line). There is no dynamo action, in spite of the high magnetic Reynolds number Rm=$10^5$.
              }
         \label{fig.6}
   \end{figure}
 There is no sign either of a small-scale fluctuation dynamo, though one may argue that we inhibit  this type of dynamo by imposing our large-scale fossil field. To check this point, we switched off the poloidal field at the latest stage of our low-$\eta$ simulation (case B). As shown in Fig.~\ref{fig.6}, the toroidal field decreases then rapidly, because it is no longer produced by the $\Omega$-mechanism, and the instability-generated field accompanies its decline. Thus the fluctuating field does not maintain itself, although the magnetic Reynolds number, Rm$\,=\,R^2 \Delta \Omega / \eta\!\sim\!10^5$ (case~B),  should amply fulfill the necessary  condition for a turbulent dynamo: at magnetic Prandtl number of order unity, as  here, the critical  magnetic Reynolds number is of order 100, according to Ponty et al. (2006). 

\section{Why do our results differ from those of Braithwaite?}

To our knowledge, the only simulation so far that claims to support a dynamo operating in stellar radiation zones  is that by Braithwaite (2006): he showed that a sufficiently strong differential rotation can amplify a seed field to a level where it seems to be maintained, while undergoing cyclic reversals.
According to him, his results confirm the analytical expectations of the role of the Pitts \& Tayler instability, but to us it is not clear whether that specific instability plays any role in his simulation: for instance the author does not mention the $m=1$ signature of the instability-generated field. Instead, he may have triggered a fluctuation dynamo. 
 
 Why do we reach different conclusions about the existence of such a dynamo? Our set-ups differ somewhat, even when we suppress the large-scale poloidal field: in Braithwaite's case, differential rotation is enforced  by a body force with strong relaxation, whereas in ours it spreads from the boundary of the computational domain, which is more realistic. 

We differ also in the boundary conditions applied to the magnetic field, which are known to play a sensitive role in numerical dynamo. We connect our internal field to a potential field outside, as if the convection zone were a perfect conductor, whereas Braithwaite imposes the field to be normal to the boundary. Using a  geometry similar to that of Braithwaite, Gellert et al. (2007) do not find dynamo action either, and they  loose  the Pitts \&Tayler instability when they switch off the exterior field, much as we find when we switch off  the poloidal field (Fig~\ref{fig.6}). 

But the main difference perhaps resides in the way the equations are solved. Our code uses (enhanced) physical diffusivities; it is of pseudo-spectral type with a resolution of 128$\times$256$\times$192, and this method is known to have exponential convergence and machine accuracy in evaluating derivatives. This allows us to reach a magnetic Reynolds number of $10^5$, and when we fail to observe dynamo action this is certainly not due to a insufficient resolution. Braithwaite uses instead a 6th order finite difference scheme, with a resolution of 64$\times$64$\times$33; the numerical diffusion is tuned to ensure stability for the chosen resolution, but it is not straightforward to infer from it the magnetic Reynolds number that characterizes the simulation (Braithwaite, private communication).

Further examining his numerical code (Nordlund \& Galsgaard 1995), one is led to guess that the background viscosity characterizing the numerical dissipation is $\nu=0.02 \, c_s \, \Delta x$, with $c_s$ being the sound speed and $\Delta x$ the grid spacing. This translates into a viscous time, which is also the Ohmic decay time, $\tau_\eta = \tau_{\nu}=
L^2/\nu= 1600 \, \tau_s$, with $\tau_s=L/c_s$.  It appears that none of the simulations reported in his paper has been run for more than $3000 \, \tau_s$; so one may wonder whether they have been carried long enough to go beyond the transient phase.

\section{Conclusion}
We have re-examined the non-axisymmetric instabilities affecting a toroidal magnetic field in a rotating star, which have first been described by Pitts and Tayler (1985) in the ideal, non-dissipative limit.  The problem was generalized by Spruit (1999) to include the diffusion of heat and of magnetic field. We have extended his analytic treatment to the case where the medium is stratified both in entropy and in chemical composition (Eq. \ref{inst-crit-general}). Our exact solutions fully validate his approximate results.  

Then we have compared these analytical results with numerical solutions built with the 3-dimensional ASH code; in our model the toroidal field is produced by shearing a fossil poloidal field through the inward propagating differential rotation imposed by the convection zone.  Our numerical solutions clearly display the Pitts \& Tayler instability with its dominant $m=1$ mode, but they do not conform to the quantitative predictions of the analytical model. In our simulations the instability occurs well below the threshold predicted by the analytical model, and it is much less sensitive to the stratification. These discrepancies are probably due to the approximations made to simplify that analytical model, such as neglecting the poloidal field, the differential rotation and the radial component of the buoyancy force. 

It also appears that the saturation of the instability cannot be ascribed to a turbulent diffusivity fulfilling the critical conditions, as in Spruit's model: it occurs when the energy of the instability-generated field reaches approximately that of the mean fields. The mean poloidal field steadily declines due to Ohmic dissipation, while it is wound up by the differential rotation to produce the toroidal field.  Contrary to Spruit's expectation,  which is based on questionable grounds as we have shown in \S\ref{section-dynamo}, we detect here no sign of a dynamo that could regenerate the mean fields; the small scale motions do not act either as an eddy diffusivity on the mean poloidal field. Unlike the turbulent motions present in a convection zone, the instability-generated motions produce here no $\alpha$ and no $\beta$-effect. Neither do we observe a fluctuation dynamo, in spite of the relatively high magnetic Reynolds number, in contrast with the findings of Braithwaite (2006),  who however considers a somewhat simpler model.

But the Pitts \& Tayler instability persists as long as the toroidal field remains of sufficient strength, i.e. a few Gauss in the conditions prevailing below the solar convection zone, which puts a similar requirement on the poloidal field. We have shown in Brun \& Zahn (2006) that such a poloidal field does not exist in the Sun, because it would imprint on the radiative interior  the differential rotation of the convection zone, and that is ruled out by the helioseismic diagnostic. The Pitts \& Tayler instability could well occur in other stars hosting a large-scale toroidal field, but we doubt that it may cause there any significant transport of matter and angular momentum, since in our simulations the motions associated with the instability behave rather as Alfv\'en waves than as turbulence. To settle that issue, observational tests will play an irreplaceable role.

\begin{acknowledgements}
Part of this work was initiated during the MSI program that was  organized in 2004 by D. Hughes, R.~Rossner and N.~O.~Weiss at the Newton Institute, Cambridge. S. M. was supported by the Swiss Science Foundation. The remarks made by our anonymous referee incited us to run an additional simulation and to clarify our differences with J. Braithwaite, who kindly provided us more information on his numerical treatment. We thank the French supercomputer centers CEA-CCRT and CNRS-IDRIS for their generous time allocations, and CNRS (Programme National de Physique Stellaire) for its financial support.
\end{acknowledgements}

\appendix

\section{Dispersion relation and energy equation}
\label{disp-en}
Let us recall that all perturbations (displacement $\vec \xi$, magnetic field $\vec b$,
pressure $P'$, temperature $T'$, molecular weight $\mu'$) are expanded in Fourier modes
\begin{equation}
\exp i(l \varpi+m\varphi+nz-\sigma t),
\end{equation}
in the vicinity of the rotation axis (the $z$-axis).

We begin by deriving the buoyancy force, which is weakened through radiative and atomic diffusion.
We split the buoyancy frequency in two parts, the first due to the thermal stratification and the second to the the composition gradient:
\begin{equation}
N^2 = N_t^2 + N_\mu^2  = {g  \over H_P} (\nabla_{\rm ad} - \nabla)
+   {g \over H_P} \left({d \ln \mu \over d \ln P}\right)
\end{equation}
with the usual notations, and taking for simplicity the perfect gas equation of state.

The linearized heat equation may be written as
\begin{equation}
- i \sigma {T' \over T} + {N_t^2 \over g} (-i \sigma \xi_z) = - \kappa s^2 {T' \over T} ,
\end{equation}
where $\kappa$ is the thermal diffusivity, and $s^2=l^2+m^2/\varpi^2+n^2$.
(We simplify the Laplacians by assuming that the diffusivities do not vary much over a meridional wavelength.)
Likewise, the advection/diffusion equation for the molecular weight perturbation takes the form
\begin{equation}
- i \sigma {\mu' \over \mu} - {N_\mu^2 \over g} (-i \sigma \xi_z) = - \lambda s^2 {\mu' \over \mu} ,
\end{equation}
with $\lambda$ being the molecular diffusivity.
Thus the buoyancy force is given by
\begin{equation}
- g {\rho' \over \rho} = g \left( {T' \over T} - {\mu' \over \mu}\right)=
- \left[{N_t^2 \over 1+i \kappa s^2/\sigma} + {N_\mu^2 \over 1+i \lambda s^2/\sigma}\right]\xi_z .
\end{equation}

Turning next to the Lorentz force,
we perturb the toroidal magnetic field ${\vec B}_t = {\vec e}_\varphi \, B  (\varpi)$ by the displacement $\vec \xi$, and draw the field perturbation $\vec b$ from the  induction equation
\begin{equation}
{\partial  {\vec B} \over \partial t}  = \vec \nabla \times ({\vec V} \times {\vec B}) -
\vec\nabla \times (\eta  \vec\nabla \times {\vec B}) ,
\label{induc}
\end{equation}
to obtain
\begin{equation}
-i \sigma {\vec b} = \vec \nabla \times (-i \sigma {\vec \xi} \times {\vec B}_t) - \eta s^2 {\vec b} .
\end{equation}
First ignoring Ohmic diffusion, we get ${\vec b} = i m\,  (B  / \varpi)\, {\vec \xi}$.
It is then straightforward to calculate the perturbation of the Lorentz force per unit volume
\begin{equation}
{\vec L} = {1 \over 4 \pi \rho} \left[ ({\vec \nabla} \times {\vec b}) \times {\vec B} 
+ ({\vec \nabla} \times {\vec B}) \times {\vec b} \right] ;
\end{equation}
introducing the Alfv\'en frequency $\omega_A^2= (B ^2 / \varpi^2)/4 \pi \rho$ and  assuming that $B  \propto \varpi^p$:
\begin{eqnarray}
L_\varpi &=& - m^2 \omega_A^2\, \xi_\varpi + (l \varpi)\, m \, \omega_A^2\,\xi_\varphi - i m \, (p+1)\,  \omega_A^2\, \xi_\varphi \nonumber \\
L_\varphi &=& i m \,(p+1)\, \omega_A^2\, \xi_\varpi \\
L_z &=& (n \varpi)\, m \,\omega_A^2 \,\xi_\varphi - m^2 \omega_A^2\, \xi_z . \nonumber
\end{eqnarray}
When keeping Ohmic diffusion ${\vec b} = i m\,  (B  / \varpi)\, {\vec \xi} /(1+ i \eta s^2/\sigma)$, and hence the Lorentz force will also be divided by $(1+ i \eta s^2/\sigma)$.

It remains to implement the expressions derived above for the buoyancy and Lorentz forces in the equations of motion:
\begin{eqnarray}
\label{eq-system}
&i& \!\!\! l  {P' \over \rho} + \left[-\sigma^2 + {m^2\omega_A^2 \over 1+i \eta s^2/ \sigma} \right] \xi_\varpi \nonumber \\
&& +\left[- {(l\varpi)m \omega_A^2 \over 1+i \eta s^2 /\sigma} + 
i { (p+1) m \omega_A^2 \over 1+i \eta s^2 /\sigma} + 2 i \Omega \sigma \right] \xi_\varphi 
 = 0 \, , \nonumber  \\
&i& \!\!\!  {m \over \varpi} {P' \over \rho} -  \left[ i {(p+1)m \omega_A^2 \over 1+i \eta s^2 /\sigma} + 2 i \Omega \sigma \right] \xi_\varpi - \sigma^2 \xi_\varphi  = 0  \, ,\nonumber \\
&i& \!\!\!  n  {P' \over \rho} - {(n \varpi) m \omega_A^2 \over 1+i \eta s^2 /\sigma} \xi_\varphi  \\
&& +\left[-\sigma^2 + {m^2\omega_A^2 \over 1+i \eta s^2/ \sigma} + {N_t^2 \over 1+i \kappa s^2 / \gamma\sigma} +  {N_\mu^2 \over 1+i \lambda s^2 / \sigma}\right] \xi_z  = 0 \, , \nonumber
\end{eqnarray}
and to complete them with the continuity equation, in the Boussinesq approximation:
\begin{equation}
l  \xi_\varpi + {m \over \varpi} \xi_\varphi + n \xi_z =0 .
\label{cont-eq}
\end{equation}
This yields a fourth-order system whose determinant is the dispersion relation we are looking for. 

From here on, we shall consider only the most realistic case where $p=1$.
To first approximation we may neglect the molecular diffusivity compared to the Ohmic diffusivity, and a fortiori to the thermal diffusivity. We further assume that $l^2 \ll n^2$, scale all frequencies and damping rates by the Alfv\'en frequency:
\begin{equation}
\widetilde\sigma={\sigma\over \omega_A} = \alpha + i \beta, \quad \widetilde\Omega={\Omega \over \omega_A}, \quad 
k={\kappa n^2 \over \gamma \omega_A}, \quad h={\eta n^2 \over \omega_A}, 
\label{define}
\end{equation}
and introduce
\begin{equation}
A_t =  {l^2 \over n^2} {N_t^2 \over \omega_A^2} , \quad 
A_\mu =  {l^2 \over n^2} {N_\mu^2 \over \omega_A^2} .
\end{equation}
The dispersion relation may then be cast into
\begin{eqnarray}
&&\widetilde\sigma^6 - \widetilde\sigma^4 [4 \widetilde\Omega^2 + A_t + A_\mu + 2 + 2 h k + h^2] 
- \widetilde\sigma^3\,  {8 \widetilde \Omega / m} \nonumber \\
&+&  \widetilde\sigma^2 \, [A_t \!+ A_\mu\! - \!3 + 2 h k \,(4 \widetilde\Omega^2 \!+A_\mu + 1)
+ h^2 (4 \widetilde\Omega^2\! + A_t+ A_\mu)] \nonumber \\
&+& \widetilde\sigma \, hk \, 8 \Omega/m - h k \,A_\mu \nonumber \\
&+& i \, \widetilde\sigma^5 \, [2 h + k]  \nonumber \\
&-& i \, \widetilde\sigma^3\, [k\,(4 \widetilde\Omega^2 +A_\mu + 2) + 2h\, (4 \widetilde\Omega^2 + A_t+ A_\mu +1) + h^2 k] \nonumber \\
&-& i \widetilde\sigma^2 \, 8 (k+h)\, {\widetilde\Omega / m} \nonumber \\
&+& i \widetilde\sigma\, [k\,(A_\mu -3) + h\,(A_t + A_\mu) + h^2 k (4 \widetilde\Omega^2 +A_\mu)] \, = \, 0 
\label{dispersion}
\end{eqnarray}

From now on we no longer deal with the gravito-inertial modes, which allows us to discard the terms in
$\widetilde\sigma^6$ and $\widetilde\sigma^5$. The slow modes then obey a 4th order equation which can be further reduced to a third order equation, by taking into account that  $4 \widetilde \Omega^2 \gg (\varepsilon A_t, A_\mu)$, $\varepsilon \ll 1$ and
 $\widetilde\sigma^2 \ll (1, k^2)$. This leads to
 \begin{eqnarray}
\varepsilon A_t  \zeta^2 &+& (2 \zeta^2+4\zeta-A_\mu) \, S^2  \nonumber \\
& -& i\, S \,\zeta \,   (\zeta^2 + 4\zeta + 3-A^* -S^2) =0 ,
\label{disp-reduced}
\end{eqnarray}
where $\zeta = 2 m \widetilde \Omega \widetilde \sigma = 2 m \Omega \, \sigma / \omega_A^2$, $\, S=2  \widetilde \Omega h = 2 \Omega \eta n^2 / \omega_A^2$ and $A^* = \varepsilon A_t + A_\mu$. 

This dispersion relation may be solved numerically to obtain the growth-rate $\sigma_I$ of the unstable modes. In the two limit cases, this equation has solutions for $S \rightarrow 0$ at given $A_\mu$ or $\varepsilon A_t$, which can be easily derived. When $A_t=0$, the growth-rate is given by
\begin{equation}
\sigma_I =  \eta \, n^2 {4\sqrt{1+A_\mu}-2-A_\mu \over 2 +2 A_\mu - 4\sqrt{1+A_\mu}} 
\quad (A_t=0, \, S \rightarrow 0);
\label{growth-amu}
\end{equation}
it is positive for $A_\mu < 6+4\sqrt3$.
In the other special case $A_\mu=0$, the growth-rate is positive between $ \varepsilon A_t=0$ and $3$: 
\begin{equation}
\sigma_I =  \eta \, n^2 {3 - \varepsilon A_t \over  \varepsilon A_t} \quad (A_\mu =0, \, S \rightarrow 0).
\label{growth-at}
\end{equation}
These expressions are not valid for $A_\mu \rightarrow 3$, or respectively $A_t \rightarrow 0, \varepsilon A_t \rightarrow 3$, where one has to keep higher order terms.

We finally turn back to the non-dissipative case, for which we derive the dynamical energy
equation. Introducing the perturbation velocity $\vec v= \partial \vec\xi /\partial t$, and considering only the real part of all variables, we rewrite (\ref{eq-system}) for the stable case $p=1$ as
\begin{eqnarray}
{\partial v_\varpi \over \partial t}&& \!\!\!\!- 2  \Omega v_\varphi + {1 \over \rho} {\partial P' \over \partial \varpi} \nonumber \\
&& + m^2 \omega_A^2 \, \xi_\varpi - m (l \varpi)\, \omega_A^2 \, \xi_\varphi 
+ 2 \, \omega_A^2 {\partial \xi_\varphi \over \partial \varphi}  = 0 \\
{\partial v_\varphi \over \partial t}&&\!\!\!\!  + 2 \Omega v_\varpi + {1 \over  \rho \varpi} {\partial P' \over  \partial \varphi}
- 2 \,\omega_A^2 {\partial \xi_\varpi  \over \partial \varphi} = 0\\
{\partial v_z \over \partial t}&& \!\!\!\! + {1 \over \rho} {\partial P' \over \partial z}  + \left(N^2 
+ m^2 \omega_A^2 \right) \xi_z - m (n \varpi) \,\omega_A^2\, \xi_\varphi = 0  .
\end{eqnarray}
We multiply these equations respectively by $v_\varpi, v_\varphi, v_z$ and add them up, making use of the continuity equation (\ref{cont-eq}), to obtain
\begin{eqnarray}
{1 \over 2}{\partial \over \partial t} \left[v_\varpi^2 + v_\varphi^2 \!\!\right. &+& \left. \!\! v_z^2 \right] +
{1 \over 2}m^2 \omega_A^2 {\partial \over \partial t} \left[ \xi_\varpi^2 + \xi_\varphi^2+ \xi_z^2 \right] 
+ {1 \over 2}N^2{\partial \over \partial t}  \xi_z^2\nonumber \\
& =&
2 \, \omega_A^2 \, \left[ {\partial \xi_\varpi  \over \partial t} {\partial \xi_\varphi \over \partial \varphi} - {\partial \xi_\varphi \over \partial t} {\partial \xi_\varpi \over \partial \varphi} \right] .
\end{eqnarray}
For these stable oscillatory modes, the r.h.s. vanishes since  
$\partial / \partial t = i \sigma$ and  $\partial / \partial \varphi = i m$; therefore the sum of kinetic and potential (magnetic + buoyancy) energies  is constant, as is the rule in non-dissipative dynamical systems:
\begin{equation}
{1 \over 2} v^2 + {1 \over 2} m^2\omega_A^2 \xi^2  + {1 \over 2} N^2 \xi_z^2 = \hbox{cst}.
\label{energy}
\end{equation}
  

\begin{thebibliography}{}

\bibitem{} Acheson, D. J. 1978, Phil. Trans. Roy. Soc. London. A, 289, 459

\bibitem{}  Arlt, R., Sule, A., \& R\"udiger, G. 2007, A\&A, 461, 301

\bibitem{}  Braithwaite, J. 2006, A\&A, 449, 451

\bibitem{} Brun, A.\ S., Browning, M., \& Toomre, J. 2005, ApJ, 629, 461

\bibitem{40} Brun, A.\ S., Miesch, M.\ S., \& Toomre, J. 2004, ApJ, 614, 1073

\bibitem{} Brun, A.\ S.,  \& Zahn, J.-P.  2006, A\&A, 457, 665 (BZ06)

\bibitem[]{} Clune, T.\ L., Elliott, J.\ R., Glatzmaier, G.\ A., Miesch, M.\ S., \& Toomre, J. 1999, Parallel Comput., 25, 361

\bibitem{} Denissenkov, P. A., \& Pinsonneault, M. 2006,  arXiv:astro-ph/0604045 (first version)

\bibitem{} Denissenkov, P. A., \& Pinsonneault, M. 2007, ApJ, 655, 1157 (final version)

\bibitem{130} Dobler, W., Stix, M. \& Brandenburg, A. 2006, ApJ, 638, 336

\bibitem{} Donati, J.-F., Forveille, T., Cameron, A. C., Barnes, J. R., Delfosse, X., Jardine, M. M. \& Valenti, J. A. 2006, Science, 311, 633

\bibitem[]{} Eggenberger, P., Maeder, A. \& Meynet, G. 2005, A\&A, 440, L9

\bibitem[]{} Gellert, M., R\"udiger, G., \& Elstner, D. 2007, arXiv:astro-ph/0705448

\bibitem{130} Goossens, M., Biront, D. \&  Tayler, R. J. 1981, Ap\&SS, 75, 521

\bibitem[]{} Gough, D. O. \& McIntyre, M. 1998, Nature, 394, 755

\bibitem{130} Heger, A., Woosley, S. E., \& Spruit, H. C. 2005, ApJ, 626, 350

\bibitem[]{} Kitchatinov, L. L. \& R\"udiger, G. 2007,  arXiv:astro-ph/0701847

\bibitem[]{} Maeder, A. \& Meynet, G. 2003, A\&A, 411, 543

\bibitem[]{} Maeder, A. \& Meynet, G. 2004, A\&A, 422, 225

\bibitem[]{} Maeder, A. \& Meynet, G. 2005, A\&A, 440, 1041

\bibitem[]{} Markey, P. \& Tayler, R. J. 1973, MNRAS, 173, 77

\bibitem[]{90} Miesch, M. S., Gilman, P. A. \& Dikpati, M. 2007, ApJS, 168, 337

\bibitem[]{} Moffatt, H. K., Magnetic Field Generation in Electrically Conducting Fluids (Cambridge Univ. Press, Cambridge)

\bibitem{} Nordlund, \AA, \& Galsgaard, K. 1995
 \\http://www.astro.ku.dk/$\sim$aake/papers/95.ps.gz

\bibitem{} Parker, E. N. 1955, ApJ, 122, 293
      
\bibitem{} Pitts, E. \& Tayler, R. J. 1985, MNRAS, 216, 139

\bibitem{} Ponty, Y., Mininni, P. D., Pinton, J.-F., Politano, H. \& Pouquet, A. 2006,
arXiv:physics/0601105

\bibitem{} Spiegel, E. A. \& Zahn, J.-P. 1992, A\&A, 265, 106

\bibitem{} Spruit, H. C. 1999, A\&A, 349, 189

\bibitem{} Spruit, H. C. 2002, A\&A, 381, 923

\bibitem{} Spruit, H. C. 2006,  arXiv:astro-ph/0607164

\bibitem{} Tayler, R. J. 1957, Proc. Phys. Soc. B, 70, 1049

\bibitem{} Tayler, R. J. 1973, MNRAS, 161, 365

\bibitem{} Wright, G. A. E. 1973, MNRAS, 162, 339
      
\end{thebibliography}
\end{document}